# A Gravitational Wave Transmitter

A. A. Jackson
Lunar and Planetary Institute, Houston
Gregory Benford
Department of Physics & Astronomy, UC Irvine

5600 words

Abstract

We consider how an advanced civilization might build a radiator to send gravitational waves signals by using small black holes. Micro black holes on the scale of centimeters but with masses of asteroids to planets are manipulated with a super advanced instrumentality, possibly with very large electromagnetic fields. The machine envisioned emits gravitational waves in the GHz frequency range. If the source to receiver distance is a characteristic length in the galaxy, up to 10000 light years, the masses involved are at least planetary in magnitude. To provide the energy for this system we posit a very advanced civilization that has a Kerr black hole at its disposal and can extract energy by way of super-radiance. Back ground gravitational radiation sets a limit on the dimensionless amplitude that can be measured at interstellar distance using a LIGO like detector.

> *My rule is there is nothing so big nor so crazy that one out of a million technological societies may not feel itself driven to do, provided it is physically possible.* --Freeman Dyson
> "The Search for Extraterrestrial Technology," 1965

This paper attempts a preliminary estimate of radiated gravitational wave signals. While this demands large sub-stellar masses moving in close orbits or high speeds, it seems possible that this emission mechanism might be used by life forms whose vast resources we cannot now envision beyond estimates, but whose signals we may register with an evolved detection technology.

The LIGO and the VIRGO detectors see black holes and neutron stars merging, using software templates derived from detailed, strong relativistic calculations. Pulling a good signal out of the vast sea of noise demands filtering from the many sources of noise. In future, gravitational wave astronomy will combat such noise problems with ever-more detailed methods to tease out even fainter signals. Perhaps, for motives we cannot well imagine, other smart beings will even encode signals in the gravitational waves that might wash through our space-time every moment.

A similar possibility emerged in the 19th century, after Maxwell predicted electromagnetic waves moving at the speed of light and Hertz, in a simple experiment using electrical circuits, detected them in radio wavelengths. Hertz thought sending signals would never happen; his waves were too weak and diffuse in spectrum, he thought. An Italian teenager heard of Hertz's remarks and thought of sending messages with the waves and made it so, along with others. That resolve by Marconi provoked a world we now enjoy. We now listen for signals from other minds across vast distances, using technologies similar to ours.

Space-time is stiff, incredibly so. Producing the slightest tremors in it demands enormous amounts of mass-energy. Even with LIGO's state-of-the-art equipment, two ordinary stars orbiting each other don't emit gravitational waves at a measurable level. To see gravitational waves demands close, grazing interactions of neutron stars or black holes.

Perhaps intelligences elsewhere now command a complex energetic instrumentality and can send possible messages to civilizations such as ours. Their motives we cannot know. Perhaps they have reason to prefer to speak to those who have mastered the far more difficult task of sensing gravitational waves, compared to the vastly simpler detection of electromagnetic signals in a myriad of possible wavelengths. Which we now have.

This study stands in the tradition of Dysonian ideas. More than a half-century ago, Freeman Dyson proposed that SETI agendas should look at technologies for harnessing an entire star's energy, building on ideas of the legendary writer Olaf Stapledon. Dyson suggested that we should look not only for signals, but for side effects like infrared emission, on scales that do not contradict physical law, but are beyond conceivable human engineering. We do similarly here for signals in gravitational waves emitted for a purpose, following on the SETI ideas evolved since the 1950s.

**Gravitational Waves and Kardashev Civilizations**

If one supposes that a civilization sends signals using gravitational waves, there are two problems to be solved: The transmitter, and the receiver. The LIGO receivers have seen gravitational radiation from natural objects. As a gravitational wave passes through matter it can change its geometry, namely a characteristic length. If one measures a length L and it responds to a gravitational wave by $\Delta L$, the 'strain' is measured by $h = \Delta L/L$. This dimensionless amplitude is very small indeed, due to the weakness of gravitational waves. LIGO can measure h to the value of $10^{-22}$, or in approximate physical terms 1/1000 the diameter of a proton.

Physically, h is related to the transmitter by $h \sim \Delta E/r$ where $\Delta E$ is a burst of gravitational radiation energy and r is the distance from the transmitter. Take $\Delta E$ as the amount of energy produced in annihilation of a mass m, namely $mc^2$, and take the distance of the transmitter to be 100 light years. The amount of energy produced can be related to the quantity of energy by a parameter specified by the Russian scientist

Nikolai Kardashev, which relates the amount of energy available to a civilization. Energy types are characterized by scales such as Type 1 'planetary', Type 2 'stellar' and Type 3 'galactic. Let K 1, 2 and 3 denote these civilizations. Table 1 shows a calculated correspondence between dimensionless amplitude, amount of energy production and civilization 'scale' for civilization located at 10,000 light years.

Table 1: Advanced civilization transmitter located at 10,000 light years

| Dimensionless Amplitude h | Mass converted to Energy (ergs) | Kardashev Scale Civilization | Gravitational Wave Receiver |
|---|---|---|---|
| $10^{-22}$ | ~0.1 Earth Mass $10^{27}$ grams | 3.6 | LIGO at 100 Hz |
| $10^{-25}$ | ~ mass of Ganymede ~$10^{26}$ grams | 3.0 | Advanced Gravitational Wave Detector ~1GHz |
| $10^{-33}$ | ~ The mass of asteroid Ida ~ $10^{17}$ grams | 2.4 | 'Planck' Length Detector |

The annihilated mass is given in grams and representative objects. LIGO can detect a Type 3 plus civilization 100 light years away, but presently only in the frequency range of ~100 Hz. A more plausible signal, we argue, may lie in the GHz range.

## 2. Gravitational Wave Machines

**2.1** A gravitational radiation machine 1

Suppose an advanced civilization makes a device to transmit information via gravitational waves. Communication means reception as well as transmission. The LIGO experiment uses an observable parameter called dimensionless amplitude, h, given by [1, 2]:

$$h = 4\frac{G}{c^2}\left(\frac{\Delta E/c^2}{r}\right) \quad (1)$$

where ΔE is the radiated gravitational energy and r is distance between observer and radiator. (G is the gravitational constant and c the speed of light.)

A fast way to estimate the radiated gravitational energy is given by [2,3,4,5] :

$$\Delta E = f(v, L)(\frac{m}{M})mc^2 \quad (2)$$

Where is m is a small mass << M, and m and M are the interacting masses. An example could be a small mass, m, falling or orbiting the big mass M. The factor f is a function of motion in the system and the and angular momentum, L, of a M. The factor f ranges between .01 for falling in, to .5 for orbiting a Schwarzschild black hole to 2.0 for orbiting a rotating black hole.

If one arrives at a circular orbit with excess kinetic energy one inserts a Lorentz factor, $\gamma = \frac{1}{\sqrt{1-\beta^2}}$, ($\beta=\frac{v}{c}$), This would be gravitational synchrotron radiation [3], with the small mass injected into an orbit about the big mass. Taking f = 1, [3,4]

$$\Delta E = \gamma^2(\frac{m}{M})mc^2 . (3)$$

To generate the radiated energy m needs to be 'deep' into the region near M's Schwarzschild radius, $r_s$. Suppose this advanced civilization has found two primordial black holes, PBH, of one Earth Mass and one tenth Earth mass. Say the Earth mass PBH is in orbit about the civilization's home star and they can maneuver the .1 earth mass black hole. (Remember we are talking *advanced civilization* here, maybe Kardashev 3.)   One model of a transmitter could be a small mass m injected into orbit about a large mass M. The geometric size of the small black holes can be computed from their Schwarzschild radius, [3]

$$r_s = \frac{2GM}{c^2} \quad (4)$$

Let M be a small black hole of one earth mass which has characteristic size of 1 centimeter , let the small mass m be of size .1 centimeters (that ~ $10^{27}$ and $10^{26}$ grams). To get the most energy out, or put another way the best amplitude at a receiver a long distance away, the small mass needs to orbit as close to M as possible. The closest distance would be at the photon sphere, $r_p$ = $3r_s$. This is the radius, for a Schwarzschild black hole of a circular photon orbit, which cannot be done by a massive particle. Now use an aspect of motion in general relativity, in which an unbound particle injected at a critical impact parameter in a region less than 10 $r_s$ will experience non-Newtonian motion. A mass on an energetic unbound trajectory injected into a close orbit about a black hole can do many orbits and then return to infinity. Aim the particle (of velocity v) at the last critical orbit, with β=v/c, the critical radius is given by, [6]

$$r_c = 2 + \frac{4}{\sqrt{8\beta^2+1}+1} \quad (5)$$

The aim point is determined by a critical impact parameter given by, [6]

$$b_c = \frac{\sqrt{8\beta^4+20\beta^2-1+\sqrt[3]{8\beta^2+1}}}{\sqrt{2}\beta^2} \quad (6)$$

These expressions are in units of the Schwarzschild radius. If the inbound orbit just shaves the critical orbit $r_c$ then it will make turns about the central mass M.

For a given injection energy the number of turns at the critical radius can be calculated. The critical impact parameter is the aim point at which the small mass goes into 'winding' orbit radiating gravitational waves. The orbits at the critical radius are unstable; after a number of turns the small mass returns to infinity. The amount of radiation can be approximated as gravitational synchrotron, use (3) to estimate it. If the impact parameter is b = $b_c$ (1+ δ) , where δ is a very small displacement. The number of orbits can be approximated by [7]

$$N \sim -\frac{\log(\delta)}{\sqrt{2\pi}} \quad (7)$$

At $r_c$ the orbital frequency is

$$\omega^2 = \frac{GM}{r_c^3} \quad (8)$$

and the frequency of the gravitational radiation is then

$$f = \frac{\omega}{\pi} \quad (9)$$

and the orbital period

$$T = \frac{2\pi}{\omega} \quad (10)$$

Take the energy radiated to be enough to excite the LIGO non-dimensional amplitude at 1000 light years. The gigaHertz frequency is higher than any current gravitational wave observatory.

The process would begin by injecting (using Kardashev 3 technology) a small mass m into an unbound orbit about M, deep in the non-Newtonian region of the black hole, executing several orbits—say, make it orbit ~10 times and return to a great distance. To keep the mechanism going, the small mass m has to be artificially returned to orbit about the big mass. The energetics are enormous. A unique feature of this system is how tiny it is! Black hole M is almost 1 centimeter and black hole m is a millimeter. The whole orbital configuration is about one meter. The civilization must array an instrumentality that can precisely aim and hit the right impact parameter, monitor and 'trim' the stability of the 'operational' orbit. The energy radiated will cause decay of the circulating orbit and the small mass m cannot be allowed to fall into the big mass black hole. The small mass must be cycled back to the big mass. The physics is allowed but the engineering physics is beyond comprehension!

Table 1 is a model of an artificial gravitational wave machine. With a given Lorentz factor γ, of 10 , that is a β = v/c of .995, a small mass injected with the critical impact parameter will make ~ 10 revolutions and radiates a pulse of ~$10^{47}$ ergs each orbit every $10^{-8}$ seconds (10 nanoseconds). The wavelength of the radiation is about 9 cm, while the exciting particle orbits at about 3 centimeters. At 1000 light years range a detector would have to be an order of magnitude more sensitive than LIGO and in a much higher frequency range.

As this model stands the orbital energy is about $10^{14}$ the solar output per second--about 1000 times the luminosity of the Galaxy!

Table 2 : Parameters for GW Machine 1

| Distance light years | 1000. |
|---|---|
| Mass1(earth mass) | 1 |
| Mass 2 (Earth Mass) | 0.1000 |
| Gamma  γ   (Lorentz Factor) | 10.00 |
| Beta    β   (v/c) | 0.9950 |
| Schwartz radius mass1 (cm ) | 0.8860 |
| Schwartz radius mass2 (cm ) | .08860 |
| Impact Parameter    ($r_s$ ) | 5.187 |
| Critical radius          (rs ) | 3.002 |
| Impact parameter delta | E-07 |
| Winding Number estimate | 10.52 |
| Kinetic Energy Incoming      (ergs) | .5377E+49 |
| Orbital Energy  -   Circular orbit    (ergs) | .4781E+48 |
| GW frequency GHz | 3.329 |
| Circular orbital period (seconds) | .1039E-08 |
| Estimated GR Energy radiated at critical orbit(ergs) | .2523E+47 |
| Dimensional Amplitude   ,h, at 1000 Ly | .8789E-23 |
| $r_s$  = Schwarzschild Radius | |

## 2.2   Gravitational radiation machine 2

Interacting body manipulation of earth mass mini black holes may be possible for a K3 civilization, but is extremely costly in energy. A second 'machine' suggests itself--an accelerated mass. Maggiore [8] calculated gravitational radiation of a mass under arbitrary acceleration, the energy ΔE radiated in a short segment of time Δt is approximately,

$$\Delta E = \frac{8G}{3c\Delta t} m^2 \gamma^2.$$

Where m is the accelerated mass and γ is the Lorentz factor. Multiply ΔE by the scaling factor $G/c^4$, and divide by distance of the source from the receiver to get the non-dimensional amplitude, h.

The adjustable parameters are the mass m, Lorentz factor γ and Δt, the source distance and an upper limit on h. If one fixes h at an Advanced LIGO value of $10^{-28}$ [8a], and the distance of the transmitter from the receiver as 10000 light years. (Note, it may be possible to measure h at the Planck Length [8b] .) This implies a 'burst' mass of 10,000 metric tons 'boosted' to a Lorentz factor γ of 1000 in an interval Δt of 1 picosecond ($10^{-12}$ seconds). This transmitter mass would have to be a mini black hole with a radius of ~ $10^{-17}$ cm and a Hawking Radiation lifetime of ~1 day. An energy of ~ 10 solar luminosities would be needed! The process would have to be repeated and modulated (in some way) to make a signal. We are back in the territory of very advanced technology, a K3 civilization.

## 3. Zoom-Whirl

We now examine the problem of what instrumentality might an advanced civilization deploy to realize a gravitational radiation machine. One guiding rule is to follow Freeman Dyson, if the physics allows it is possibly technologically feasible. That does not mean a civilization will marshal the economic or sociological forces to realize an extreme artifact. Mastering the energy requirement for this kind of machine is extreme. But physics, as we now fathom it, defines the difficulties.

A civilization orbiting a solar type star at best has ~$10^{33}$ ergs/sec on tap. (Other power sources will probably be less.) Set some parameters and see what kind of transmitter might be built. First set the sensitivity of a 'LIGO' like receiver. Physical considerations seem to imply one could measure the dimensionless 'strain', h, as $10^{-29}$ (private communication), h is approximately $10^{-22}$ at the present. Take as a characteristic distance to an advanced civilization in our galaxy as 10,000 light years.

Suppose the civilization uses the mechanism in section 2--a small mass flies by a larger mass at a very close distance. From equation 3 the energy in a pulse is a function of the mass ratio and the mass of the flyby 'exciter'. To transmit over interstellar distances the exciter should be of substantial mass, some fraction of an earth mass, which must be maneuvered in orbit. The exciter mass is injected into a close-encounter orbit where it 'winds' for a finite number of revolutions (whirls) and then returns to a large distance. Changing the orbit of the exciter will require substantial energy. A trial and error calculation using the modeling in section 2 shows that if the primary mass is one Earth mass then a $10^{-4}$ Earth mass exciter (about the mass of Ceres) injected with a Lorentz factor of 2 can excite a detector (h~$10^{-29}$) at 10,000 light years. Take the central mass as a mini-black-hole of radius ~1cm while the exciter has a radius of ~ 1 micron.

A scenario could be as follows. A Kardashev 2 civilization has harvested primordial sub-stellar mass black holes. An earth mass black hole is placed in orbit some distance away from the home planet; call this the 'primary' of the gravitational machine. Then a smaller mass, the 'exciter', mass $10^{-4}$ of the primary, gets injected in with excess energy (that is, an unbound orbit). It shaves the critical radius at ~5 Schwarzschild radii ($r_s$). At this radius a mass can wind for a finite number of orbits; below this radius it plunges into the primary. While riding this knife edge orbit, the exciter radiates energy, approximately $10^{41}$ ergs per pulse. While this orbit shaves the critical radius by approximately $10^{-6}$ of its radius, the total energy radiated is $10^{-4}$ the orbital energy. After ~ 10 orbits it returns to 'infinity'—back to the cycling processor. To repeat the process, the civilization must turn the exciter mass such that it has the same impact parameter as before, and so repeats the process.

It may be possible to configure the operational orbit such that is periodic (see appendix). With the periapsis and energy arranged, just so, one can obtain repeating zoom-whirl orbits (see below). Note not only to get the non-Newtonian motion, the exciter mass needs to pass close to the primary, since the pulse generated goes like $1/r^5$. The configuration is described in this table:

Table 3. The Orbital Gravitational Machine

| Parameters | Value | Units |
|---|---|---|
| Distance | 10000 | Light Years |
| Central Mass | 1 | Earth Mass |
| Exciter Mass | .1 | Earth Mass |
| Lorentz factor γ | 2 | |
| Percent the speed of light β | .866 | C |

| Schwarzschild radius central mass M | ~1 | Cm |
| :---: | :---: | :---: |
| Schwarzschild radius exciter mass m | ~.1 | Cm |
| Critical Impact Parameter $b_c$ | 5.7 | Schwarzschild radii of M |
| Critical Impact radius | 5.0 | Schwarzschild radii M |
| Aim point δ (for [$b_c$ (1+δ)]) | $10^{-6}$ | |
| Number of turns at the critical radius | ~9.35 | |
| Orbital period at the critical radius(~$b_c$) | $10^{-7}$ | Seconds |
| Incoming Energy | $10^{49}$ | Ergs |
| Approximate orbital energy at circular radius | $10^{48}$ | Ergs |
| Gravitational energy radiated per winding orbit | $10^{47}$ | Ergs |
| Frequency of the radiation | 3.3 | GHz |
| Note: The sun radiates | $10^{33}$ | ergs/sec |

Generally, injecting the small mass into an eccentric orbit around a black hole means those orbits decay in a flash. However, a mass injected with kinetic energy can skim the knife edge at the photon sphere, ~3 $R_s$ where $R_s$ is a Schwartzschild radius.. This can still be an open orbit--that is, returns to infinity—thus allowing a large region where the charged mass can be energized again, with velocity and spin added electromagnetically in an accelerator. These tiny objects on the scale of centimeters can be controlled electromagnetically, through their charges and magnetic moments. A small mass m with the right impact parameter can come in from infinity and can go into winding orbits that circle at 3m about 10 times and then escape. The same is true of ultra-relativistic trajectories of objects with mass. Relativistic orbits very close to black holes precess around the black hole and form nested sets.

The picture below was produced using the equations of the online site *Relativity 4 Engineers*, plotted with a new color for every full orbit. (http://www.einsteins-theory-of-relativity-4engineers.com/relativistic-orbits.html) The orbit 'whirls' around the black hole a few times and then 'zooms' out to the apoapsis and back again - a so-called *whirl-zoom* orbit. This happens when the periapsis is very close to the black hole, between two to three times the event horizon radius. If the orbiting particle comes much closer, it will either fall into the black hole, or it will escape completely, depending on the total orbital energy of the particle.

Making this happen demands delicate control of the orbit and energy, using methods we do not now know. The central hole's spin adds to the incoming hole's angular momentum, boosting the orbit back out. Zoom-whirl behavior is characteristic of strong relativity and radiates harmonics in the gravitational waves—the key to imposing high-bit-rate signals on the outgoing gravwave train.

Figure 1 Nonlinear zoom-whirl orbits, color coded for each pass.

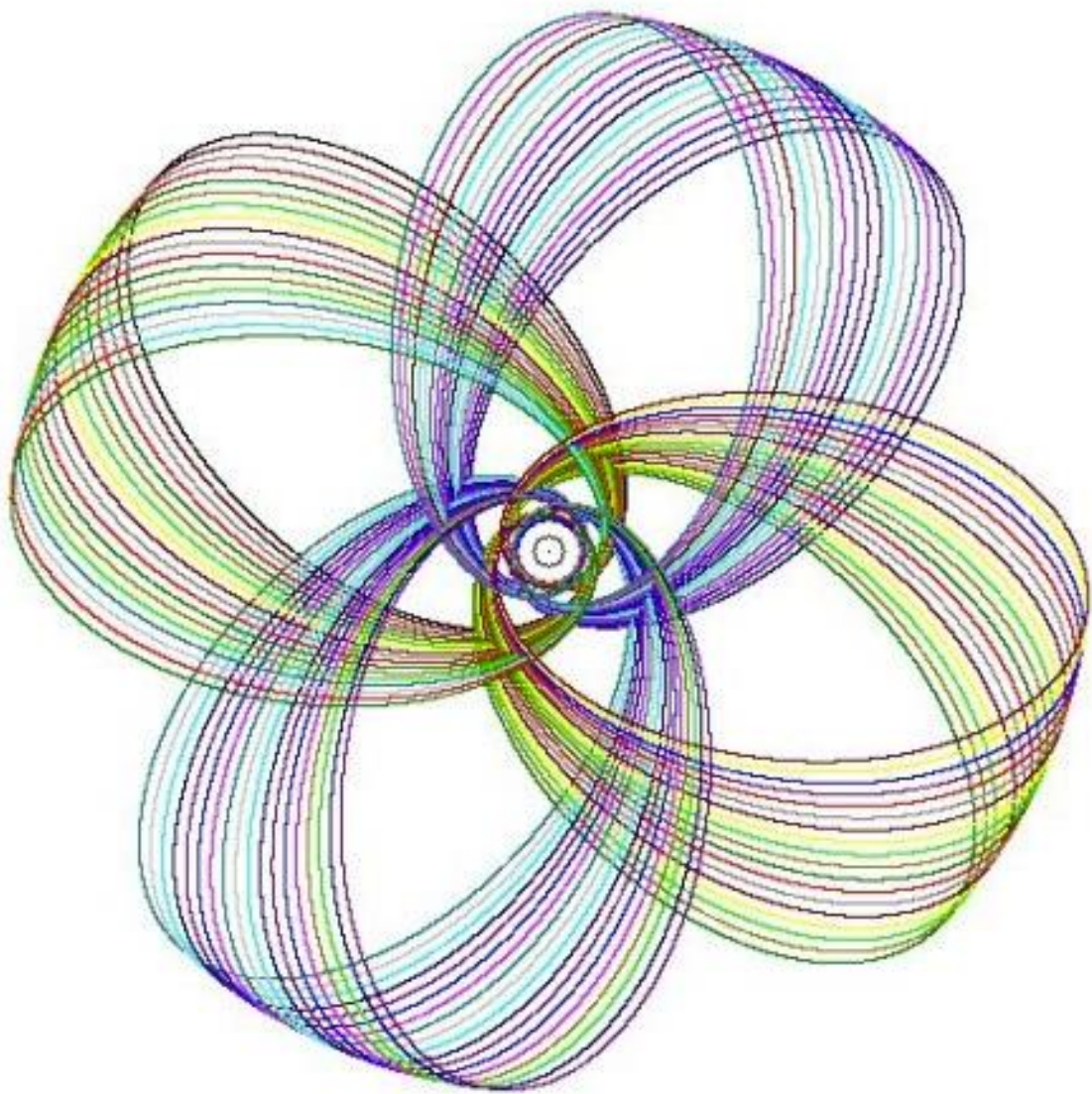

As an example, consider a 0.1 Earth mass m forced into an orbit around a 1.0 Earth mass central body, M, both black holes of size less than a centimeter. The smaller mass m is injected into a nearly radial trajectory toward the central mass M. With a suitable angular momentum embedded into the smaller m, the mass twists into a class of zoom-whirl orbits, approaching no closer than 5 centimeters, i.e., well beyond the critical radius from which it cannot be recovered, ~ 3 cm. The orbital kinetic energy of this $5.4 \times 10^{48}$ erg/sec and potential energy of $4.8 \times 10^{48}$ erg/sec.
The zoom-whirl orbits emit gravitational radiation at frequency ~1 GHz while executing orbits in ~ $10^{-7}$ sec.
The zoom-whirl orbits last through an estimated number of loops made before escape of about 13.5--full orbits before liberating, the winding number. The total emission interval is a microsecond. This means at least a thousand bits of information can be emitted before the smaller mass returns to the energy source to be resupplied. The gravitational synchrotron loss is just less than kinetic energy. The civilization must make sure that m does not crash into M, so m flying out would have to be deflected by the expenditure of energy, re-accelerated and sent back in again. This accelerator can be large, but the emitting staccato 'ticker' is quite small, only centimeters in size. One might envision a number of small masses as the 'exciters' orbiting in a complicated convoy to make a more intense, signal-rich ensemble. These can then emit in concert, insuring a longer message.
Waveforms are modulated by the harmonics of zoom-whirls, showing quiet phases during zooms and louder glitches during whirls. Zoom-whirl behavior in spinning pairs is a common feature of eccentric orbits, despite the drain of gravitational radiation. Fine tuning of initial conditions at the outer limit of the orbits, apastron, imposes the emitted message. (See Figure 2 for an example gotten from a detailed numerical simulation using full general relativity.)
**Figure 2** Characteristic shape of the waveform in a zoom-whirl orbit.

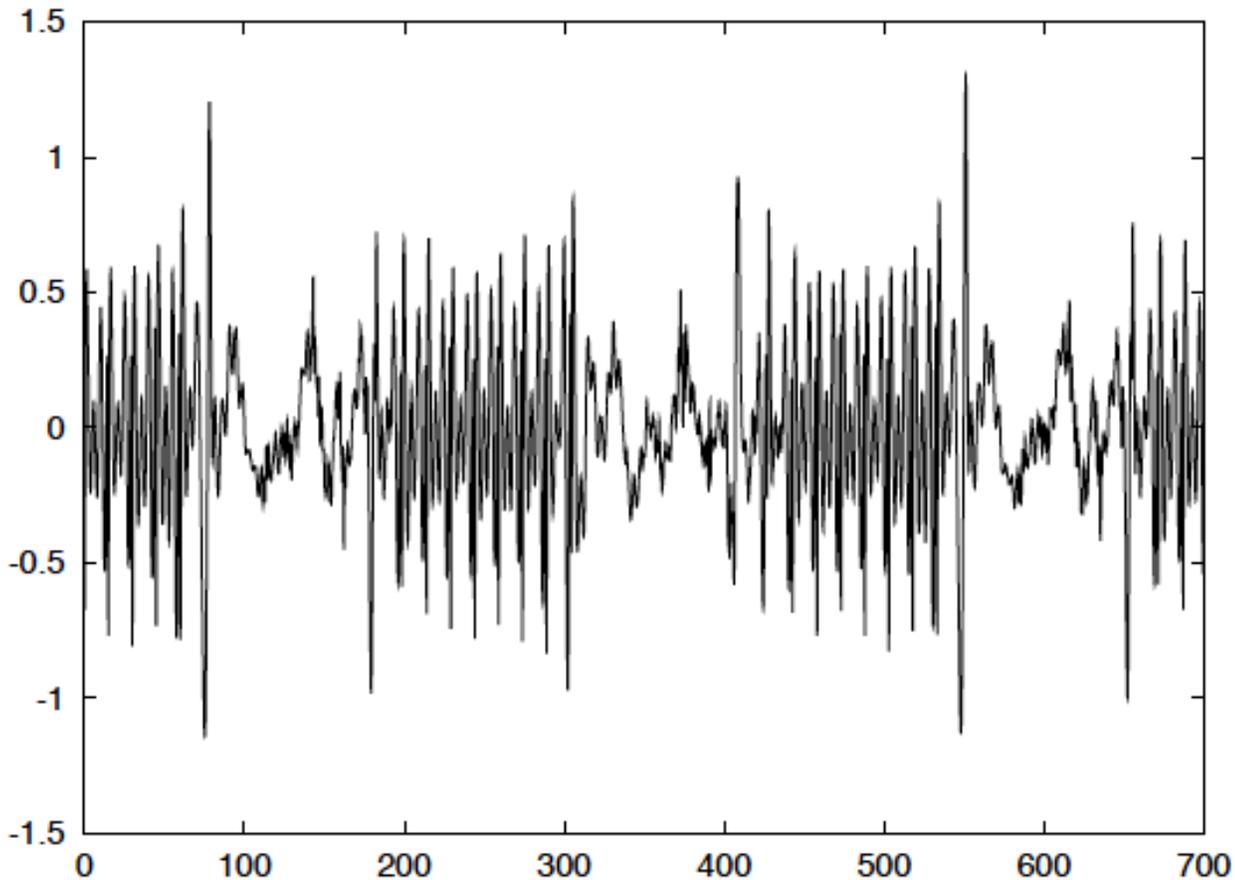

The orbit emits a periodic succession of high-amplitude/high-frequency parts (coming from the whirling motion of the particle near the periastron) and intervening low-amplitude/low frequency parts (from the zooming in and out motion, apastron). Quadrupole ($\ell = 2$) emission dominates gravwaves.[5]

     Of course, managing the black hole and gravwave stresses in the central region of the emitter is crucial. The emitter sits the core of a relatively low-mass surrounding structure that must flex and endure severe stresses. The wave stretches one transverse direction while the other compresses. Because of potential accident, the entire assembly should orbit well beyond any inhabited zones.

The total gravwave energy emitted in this example is about a hundredth of an Earth mass. How can this be replenished? Obviously, a culture able to handle such masses and thus energies must have sources we do not know. The precursor to any such project must be using black holes to extract energy. Such energy sources might be essential in mining operations throughout an outer solar system, where large, sporadic energies are useful in mass processing for industrial use.

One notes that besides giving the exciter mass a kinetic energy of $10^{49}$ ergs, an orbital energy of $10^{48}$ must also be attained and modified. These energies are $10^{16}$ greater than a local solar type star output. Where would this energy come from?

**4.0 The Kerr Bomb**

To make such energies, now suppose the advanced civilization has 3 small black holes in its inventory. Two are the 'orbital machine', the larger central mass plus the exciter mass. The machine central and exciter black holes form a binary system orbiting the home star.

The third black hole is a rotating (Kerr) black hole--the 'machine' power plant. It exploits super-radiance [10], to extract energy from the rotating black hole by scattering electromagnetic radiation from it. Radiant Energy from the home star is diverted to the black hole bomb, where it gets amplified by superradiance. Surround the power house black hole with a spherical mirror about 300 meters in diameter [11]. Figure 3 shows a cut away of the 'bomb transmitter-mirror' system. This scattered radiation will pick up a small amount of energy from the rotating event horizon. If the scattered radiation is then confined, for merely a short length of time, by a spherical 'mirror,' then an enormous amount of energy may be made available. Suppose the 'power-house' black hole has a mass of Jupiter and the Kerr black hole is rotating at its maximum. This means the horizon would be spinning at ~ 1 GHz , as long as the impinging radiation has a frequency of less than or of the order of this it will be amplified. One notes an immediate problem with the mirror, if the radiation is to be contained to a level $10^{16}$ times bigger than the sun's output then the mirror would have to have a tensile strength greater than neutron star material!

Figure 3 The Kerr power plant, a pulsating black hole bomb.

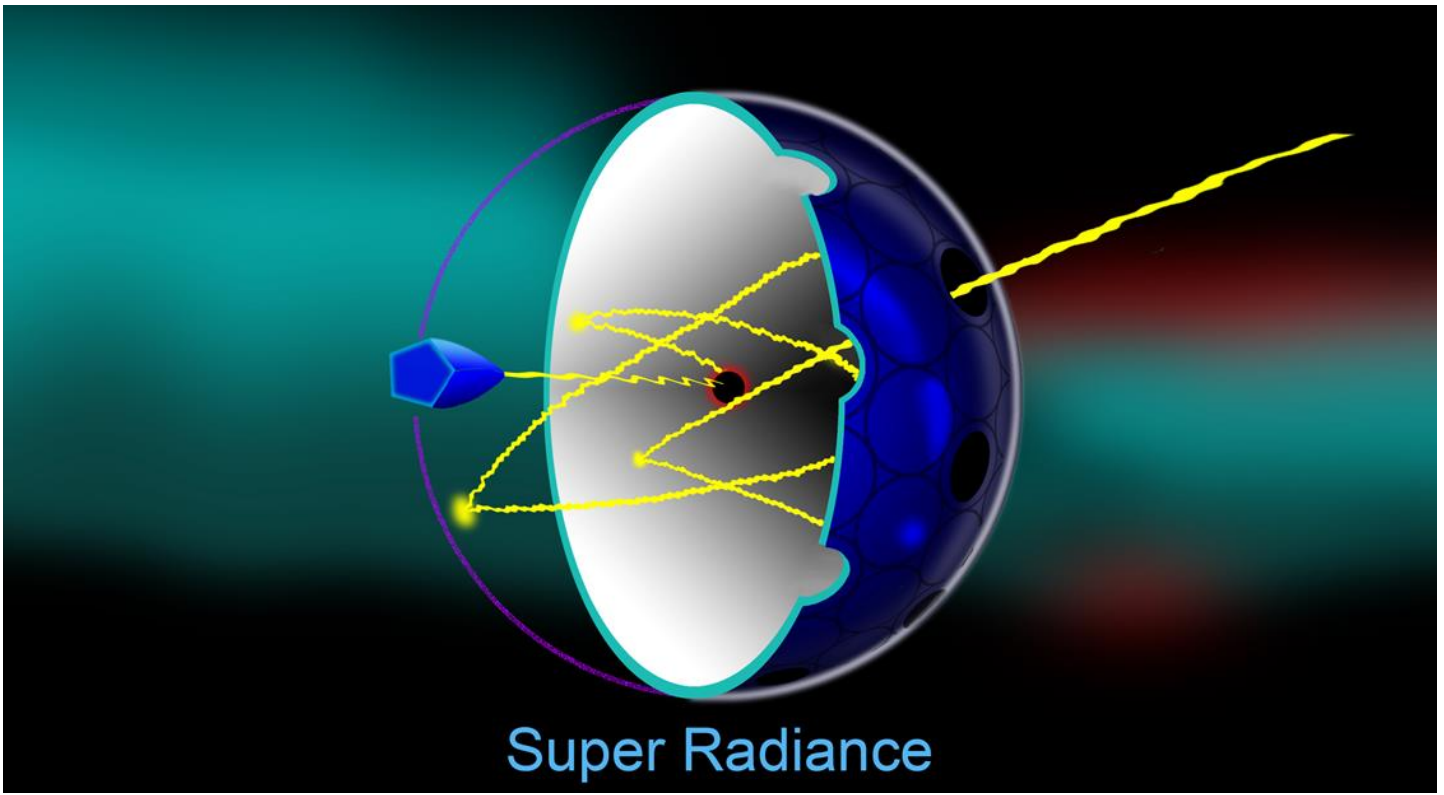

Figure 4 is a schematic representation of total 'gravitational wave machine' system. Note the 'Superradiance machine' is inward of the gravitational wave machines; the Home Star is located somewhere in this system.

Figure 4 : Schematic of an orbital gravitational machine.

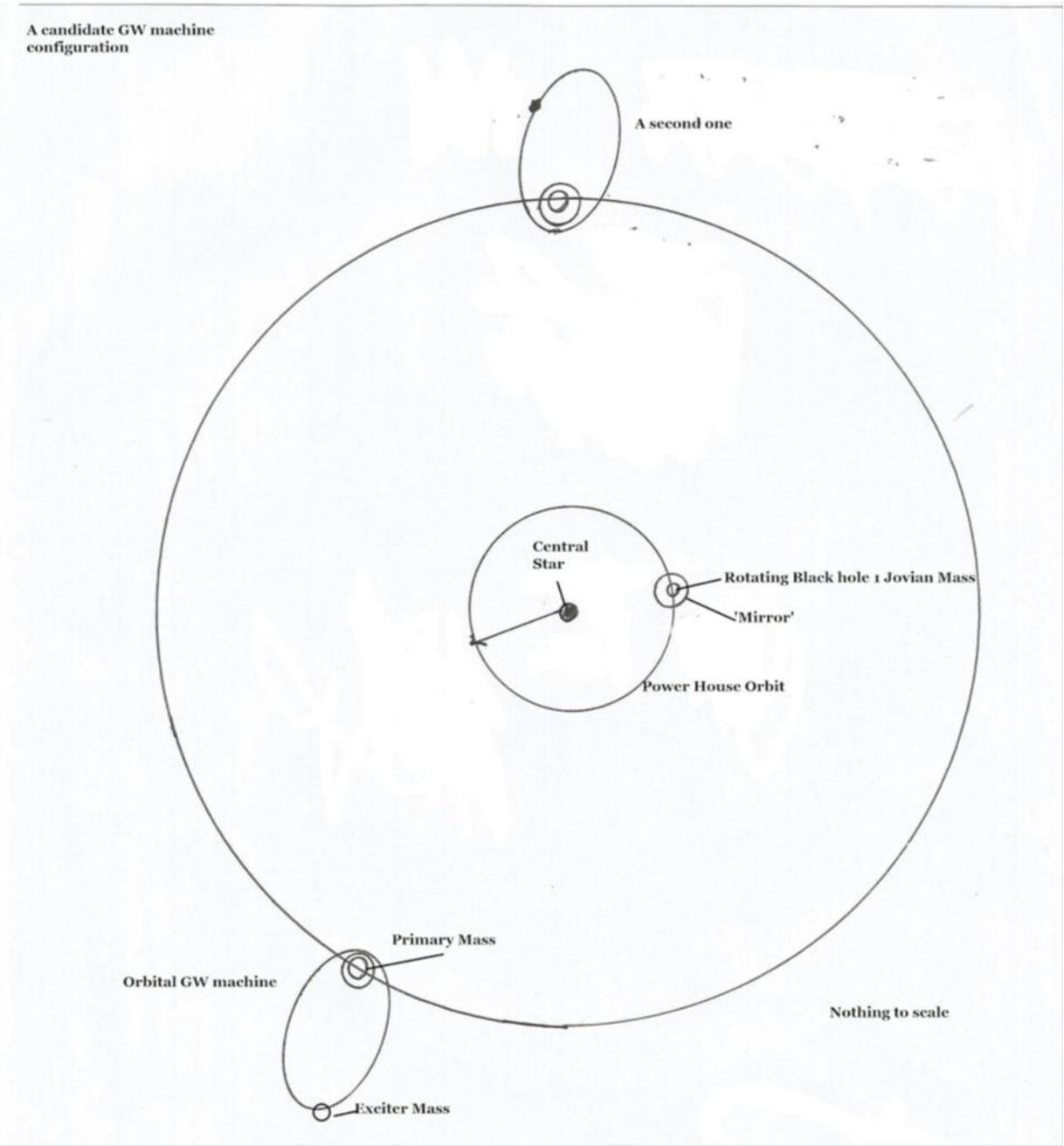

If a Kardashev civilization of order 2 has mastered the energy output of its local star, that makes some of this available (about .001 of its local star's output) to transmitters located on it's local Kerr black mirror. Then in ~ 1 second (the e-folding time) there is an amplification of $10^{17}$, producing the energy needed to manipulate the orbit of the exciter mass. The extra energy extracted from rotational energy of the Kerr black hole can be used to maintain the orbital machine's configuration. Arrange the 'bomb' such that it pulses--that is, before the sphere can blow up the energy is transferred to the gravitational machine. For a sphere only hundreds of meter's to a kilometer in diameter one seemingly could have an almost solid sphere whose stability would have to be controlled. This Kerr power plant could supply the energy needed for both GW machines mentioned above. There is a problem in that making the mechanics of the engine to work one is generating almost a supernova power output of electromagnetic energy. Containment of this energy could mean in the end product one has a 'waste heat' object of extraordinary brightness!

How such an instrumentality could be realized is mind boggling. A naïve estimate shows that the mirror has to be made of something strong, about 15 orders of magnitude greater than the tensile strength of Graphene—actually, more than the tensile strength of neutron star matter. So these demands stretch the credibility of any Dysonian engineer. But other tricks to offset such pressure may arise from superior engineering methods. Recall that buildings readily constructed now were thought impossible only a century ago.

## 5. Electromagnetic leakage

We must mention that for a gravwave-emitting society, not letting the electromagnetic energy radiate to make its own beacon may be essential.

The varying EM fields will be plausibly of lower frequency. They are not changing the zoon-whirl orbits, they're powering the slower accelerations, so may be an order or magnitude lower in frequency, to 100 MHz or so, but powerful. So then the emitters could make such leakage uninteresting, by other methods.

One idea is to capture the electromagnetic energy further away, with a reflecting metallic sphere. Can this be the same sphere in the Kerr bomb? Best to make it a perfect conductor, feeding energy back in, or absorbed/routed and saved for accelerating the next black hole to high energy. Capturing reflectors which store the electromagnetic energy until needed, by routing it through waveguides and resonant storage cavities, can shape the next, needed accelerating electromagnetic energy wave that drives the black holes up to the needed energy. Recycling such vast energies would be essential. Many other tricks to reuse photons may arise from superior engineering methods.

A further way to avoid seeming like an electromagnetic signal would be to deliberately blur the leaked emission. Phase-delaying, merging and time-staging leakage strips it of signal. This makes even a bright emission seem like an astronomical oddity, rich in energy but not in meaning.

## 6. Conclusions:

The great equalizer in all communication across vast ranges is the speed of light and gravwaves alike. This leads to motives, once societies achieve technologies. The grand goals of alien minds can be imagined [12]. Briefly, they can be

- *Kilroy Was Here*—memorials to dying societies.

- *High Church*—records of a culture's highest achievements. The essential message is *this was the best we did; remember it.* A society that is stable over thousands of years may invest resources in either of these paths. The human prospect has advanced enormously in only a few centuries; the lifespan in the advanced societies has risen by 50% in each of the last two centuries. Living longer, we contemplate longer legacies. Time capsules and ever-proliferating monuments testify to our urge to leave behind tributes or works in concrete ways. The urge to propagate culture quite probably will be a universal aspect of intelligent, technological, mortal species (Minsky, 1985).

- *The Funeral Pyre:* A civilization near the end of its life announces its existence.

- *Ozymandias*: Here the motivation is sheer pride; the Beacon announces the existence of a high civilization, even though it may be extinct, and the Beacon tended by robots. T

- *Help!* Quite possibly societies that plan over time scales ~1000 years will foresee physical problems and wish to discover if others have surmounted them. An example is a civilization whose star is warming (as ours is), which may wish to move their planet outward with gravitational tugs. Many others are possible.

- *Join Us*: Religion may be a galactic commonplace; after all, it is here. Seeking converts is common, too, and electromagnetic preaching fits a frequent meme.

These motives may well persist into cultures vastly more powerful than ours, who prefer gravwave signals to the easier electromagnetic ones. They may choose gravwaves because they do not wish to be known to mere electromagnetic civilizations. LIGO has opened a window that perhaps few societies in our galaxy could manage, or wish to. It may show us more than astronomers expect.

It is also possible that we have overestimated the technical difficulties. Received power can be enhanced in transmissions if sources are made coherent. In gravitational radiation, this would mean resonant paralleling of trajectories in the smaller masses, as they orbit the central mass. This might be possible, but will greatly complicate matters, as the zoom-whirl orbits are already highly nonlinear; adding a coherence constraint makes their management more difficult. One can also imagine an array of more than one m/M system. This would mean spacing the large mass elements in order for their emissions to align. Such an array then can direct emissions in a narrower spatial and perhaps frequency band, just as in electromagnetic systems. Some efficiency improvements seem possible this way. . If so, the threshold of gravwave emitters may be low enough to make it a commonplace of truly long-lived societies.

## Acknowledgments

We are grateful for discussions with Martin Rees, James Benford, Freeman Dyson, and Richard Matzner.